\documentclass{webofc}
\usepackage[varg]{txfonts}   
\begin{document}
\title{SQM2022: Theoretical Summary}
%

\author{\firstname{Berndt}\lastname{M\"uller}\inst{1,2}\fnsep\thanks{\email{mueller@phy.duke.edu}} 
}

\institute{Department of Physics, Duke University, Durham, NC 27705, USA
\and
	Wright Laboratory, Yale University, New Haven, CT 06520, USA
          }

\abstract{%
Written version of the theoretical summary lecture presented at the Strangeness in Quark Matter 2022 conference.
}
\maketitle
\section{Introduction}
\label{intro}

In his Theoretical Overview, J.~Kapusta \cite{Kapusta} described a large array of topics and open questions that fall within the purview of this conference series. In my summary lecture I will focus on a reduced subset of topics where I believe significant progress was presented at SQM2022. I apologize in advance to those whose beautiful work I have no time and space to cover. The absence of a mention does not reflect negatively on their contributions. Above all, it is worth emphasizing the following points:
\begin{itemize}
\item Remarkable progress continues to be made.
\item Unexpected new insights continue to be generated.
\item The topics of “SQM” have expanded far beyond what was imaginable 40 years ago.
\item The huge amount of data of a wide variety being generated by the experiments is both, a benefit and a curse.
\item It is vital that the community stays focused on its core goals.
\end{itemize}
My top-level assessment of the status of the field is:
\begin{itemize}
\item We have an accepted ``standard'' reaction model for the central rapidity region in collisions at $\sqrt{s_\textrm{NN}} > 100$ GeV.
\item We do not have a widely accepted and data tested model for collisions at $\sqrt{s_\textrm{NN}} < 20$ GeV.
\item Theoretical tools for the baryon-rich region remain inadequate.
\item The mechanisms for hadron emission remain insufficienty understood.
\item Probes of chiral symmetry restoration remain at best tenuous.
\end{itemize}
In the following, I will discuss newly discovered possible flaws in the standard reaction model, core-corona models, the physics of collisions at the energies of the RHIC beam energy scan, critical fluctuations, anomalous chiral phenomena, hadron emission, heavy quarkonia, strangeness, and Bayesian model-to-data comparison.

\section{Reaction Dynamics}
\label{sec-1}

The standard reaction model for high-energy collisions relies on a multi-stage approach beginning with some model of the initial energy deposition and followed by, in sequence, hydrodynamic expansion of the quark-gluon plasma (QGP) fluid, Cooper-Frye hadronization, and hadronic Boltzmann transport. All these stages need to be carefully checked for intrinsic consistency and for consistency of the transition between them. Many careful investigations have addressed this issue in the past and validated important aspects of the framework. 

At this conference, C.~Plumberg \cite{Plumberg} reported recent work that takes a careful look at the validity of using second-order viscous relativistic hydrodynamics during the early phase of the hydrodynamic stage. He applied a set of criteria to ascertain whether the evolution of all fluid cells is causal or not. Causality requires that the propagation velocity $v$ satisfies the condition $0 \leq v^2 \leq c^2$, which means that the system of differential equations is hyperbolic and propagation is not superluminal. These requirements result in 6 necessary and 8 sufficient conditions that can be checked for each fluid cell. 

The implementation he studied involved energy deposition in the IP-glasma model, possibly followed by a period of free streaming or kinetic theory transport. What he found for central Pb+Pb collisions at $\sqrt{\textrm{NN}} = 2.76$ TeV is remarkable: (a) When fluid dynamics is initiated at $\tau = 0.4$ fm/c directly from the IP-glasma initial conditions, almost all fluid cells strongly violate causality. (b) When fluid dynamics is preceded either by free-streaming or kinetic transport and initialized at $\tau = 0.8$ fm/c, the necessary, but not the sufficient, conditions for causality are met over a large part of the volume, but causality remains strongly violated near the surface. The violation of causality persists over a period of approximately 20\% of the duration of the hydrodynamic stage, during which more than half of the final elliptic flow is generated.

It is currently unclear what a suitable remedy might be that retains the indispensable features of the fluctuating nature of the initial state. The causality violations are associated with the large initial density fluctuations that drive higher Fourier components of the collective flow. Is it perhaps possible to work directly with moments of the initial density fluctuations instead of the full fluctuating density distribution by performing a Fourier decomposition of the fluid in azimuthal angle?

The strong causality violations near the outer edge of the fireball indicate that it is preferable to treat the core and the corona of the fireball differently, applying fluid dynamics to the core and some form of kinetic transport to the corona.  In her talk Y.~Kanakubo \cite{Kanakubo} presented a dynamical core-corona model that allows quasiparticles (partons) in the corona to thermalize and become part of the core. The results appear encouraging: While the overwhelming part of the fireball joins the fluid core in Pb+Pb collisions, the corona constitutes most of the fireball in p+p collisions, and even in the most central p+p collisions approximately half of the fireball remains in the corona phase. The transition to core dominance occurs at $dN_\textrm{ch}/d\eta \approx 20$ for both systems. Some interesting questions arise: Can core-corona models of this type for midrapidity be formulated as a rigorous effective kinetic theory of QCD similar to the kinetic equilibration scenarios? What would be the limits for such a formulation? Would a kinetic theory avoid causality violations and provide a more realistic description of the entire mid-rapidity fireball?

While energy deposition at high energies can be described as an instantaneous quench owing to the high degree of Lorentz contraction of the two colliding nuclei, auch an approach fails at the lower energies of the RHIC beam energy scan. In the target rest frame a modified form of this picture may survive because the projectile nucleus is still highly Lorentz contracted to $2R/\gamma_{\rm Lab}$. The collision then ptakes the form of a time-staggered local quench as the sheet-like projectile sweeps over the target nucleus, and a space-like Cauchy hypersurface for hydrodynamics initial conditions $t_i(z)$ can still be defined? Does this allow to develop a modified ``standard model'' for the reaction? 

In the center-of-mass frame on the other hand, both nuclei are only mildly Lorentz contracted to thickness $2R/\gamma_\textrm{cm}$, and the collision cannot be treated as a quench because energy and baryon number deposition occur over extended period of time.  As T.~Mendenhall \cite{Mendenhall} showed in his talk, this limits the attainable maximum energy and baryon number density. For example for a central Au+Au collision at $\sqrt{s_\textrm{NN}} = 5$ GeV the maximum energy density of slightly more than 1 GeV/fm$^3$ and baryon number density $n_B \approx 0.7$ fm$^{-3}$ is reached only at $\tau \approx 4.5$ fm/c. This implies that the fireball conditions $(T,\mu_B)$ at the time of highest density are rather close to the QCD phase boundary when also a formation time $\tau_\textrm{F}\approx 1$ fm/c is taken into account.

\section{Critical point and anomalous chiral phenomena}
\label{sec-2}

An extension of the new parametrization of the lattice equation of state at non-zero baryon chemical potential to nonvanishing strangeness chemical potential was presented by P.~Parotto \cite{Parotto}. The new parametrization produces an {\it ab initio} equation of state for conditions prevailing in heavy ion collisions (strangeness neutrality) for $\mu_B/T \leq 3.5$ with reasonable uncertainties and covers a large part of the QCD phase diagram explored in the RHIC beam energy scan (down to $\sqrt{s_{\rm NN}} \approx 6$ GeV).

Much theoretical progress has been made on modeling the dynamics near the conjectured critical point in the QCD phase diagram. In her talk at SQM2022 M.~Pradeep \cite{Pradeep} explored the off-equilibrium critical dynamics by treating the critical mode as an independent degree of freedom in the framework of fluid dynamics (the ``Hydro+'' approach \cite{Stephanov:2017ghc}). The delayed response of the critical mode avoids the singularity that occurs at the critical point but leads to enhanced remnant baryon number fluctuations at freeze-out. For reasonable parameter values the remnant fluctuations are found to depend only weakly on the freeze-out temperature.
  
In his talk on anomalous chiral phenomena E.~Finch \cite{Finch} discussed the results from the RHIC isobar  ($^{96}$Zr -- $^{96}$Ru) comparison run \cite{STAR:2021mii}. The results, which were obtained by means of a blind analysis, have been re-assessed for various isobar specific corrections that were not understood at the time of the analysis. The STAR collaboration has concluded that these reduce the expected ratio (Ru/Zr) for various putative signatures of the chiral magnetic  effect (CME) by approximately 3\% from unity to 0.97.  The measured values for all signature CME observables agree with this revised baseline within 0.5\%.

Efforts to search for other manifestations of anomalous chiral phenomena will surely continue. However, I believe that it is now time for theorists to embark on a rigorous analysis what these impressive results mean. What are the implications of the absence of CME signatures at detectable levels? There are three main drivers of the magnitude of the CME observables: (1) the degree of chiral symmetry restoration in the QGP, (b) the size of QCD winding number fluctuations in the QGP, and (c) the magnitude of the remnant magnetic field in the QGP at late times. Theorists will need to ask which of these quantities are constrained by other observables. 

It seems to me that the lattice provides compelling evidence of chiral symmetry restoration even in the absence of direct experimental confirmation. We have experimental bounds on the remnant magnetic field at hadronization from hyperon polarization data \cite{STAR:2020xbm} which are not far from state-of-the-art theoretical predictions \cite{Grayson:2022asf}. If this assessment is correct, the experimental data can set an upper bound on winding number fluctuations over the course of the heavy ion collision. How valuable would this information be? Given the tremendous effort invested in the isobar experiment a commensurate theoretical effort is called for.

\section{Hadronization}
\label{sec-3}

Because the dynamics of QGP hadronization is neither described perturbatively nor by lattice simulations, it is among the largest unknowns in models of relativistic heavy ion collisions. W.~Zhao \cite{Zhao} discussed a phenomenological framework that treats bulk fluid hadronization, quark coalescence, and fragmentation consistently within an end-to-end dynamical model of the  collision. He concluded that data from Pb+Pb collisions at LHC and Au+Au collisions at RHIC both require the presence of all three hadronization mechanisms in order to reproduce the measured valence quark number scaling. A similar conclusion was reached by V.~Minissale \cite{Minissale} who presented work by the Catania group on $D^0$ and $\Lambda_c$ production as measured by STAR and ALICE. He suggested that recombination may even be required to describe the $\Lambda_c/D^0$ ratio in p+p collisions at LHC for $p_T<5$ GeV/c.

Quite a few talks discussed new data that probe aspects of the statistical hadronization model (SHM). C.~Pinto \cite{Pinto} showed that the SHM describes particle yields in Pb+Pb collisions at $\sqrt{s_{\rm NN}} = 2.76$ GeV over 9 orders of magnitude from light hadrons to alpha particles with a common formation temperature $T_{\rm chem} \approx 156$ MeV. This temperature coincides remarkably well with the pseudocritical temperature $T_c$ ``measured'' on the lattice. He pointed out that the hypertriton $^3_\Lambda$H with its intrinsic radius $a > 5$ fm provides for an extreme test case. 

The common expectation is that $^3_\Lambda{\rm H}/\Lambda$ yield ratios for SHM only vary with system size because of exact strangeness neutrality of the fireball. In a coalescence scenario the ratio depends on the size of the $^3_\Lambda$H bound state and falls off precipitously with shrinking fireball size. Recent data for this ratio from p+Pb and p+p collisions clearly favor two-body coalescence -- a reasonable assumption since the  $^3_\Lambda$H can be thought of as a loose d+$\Lambda$ bound state -- over the SMH prediction with strangeness neutrality. 

\begin{figure}[h]
\centering
\includegraphics[height=3cm]{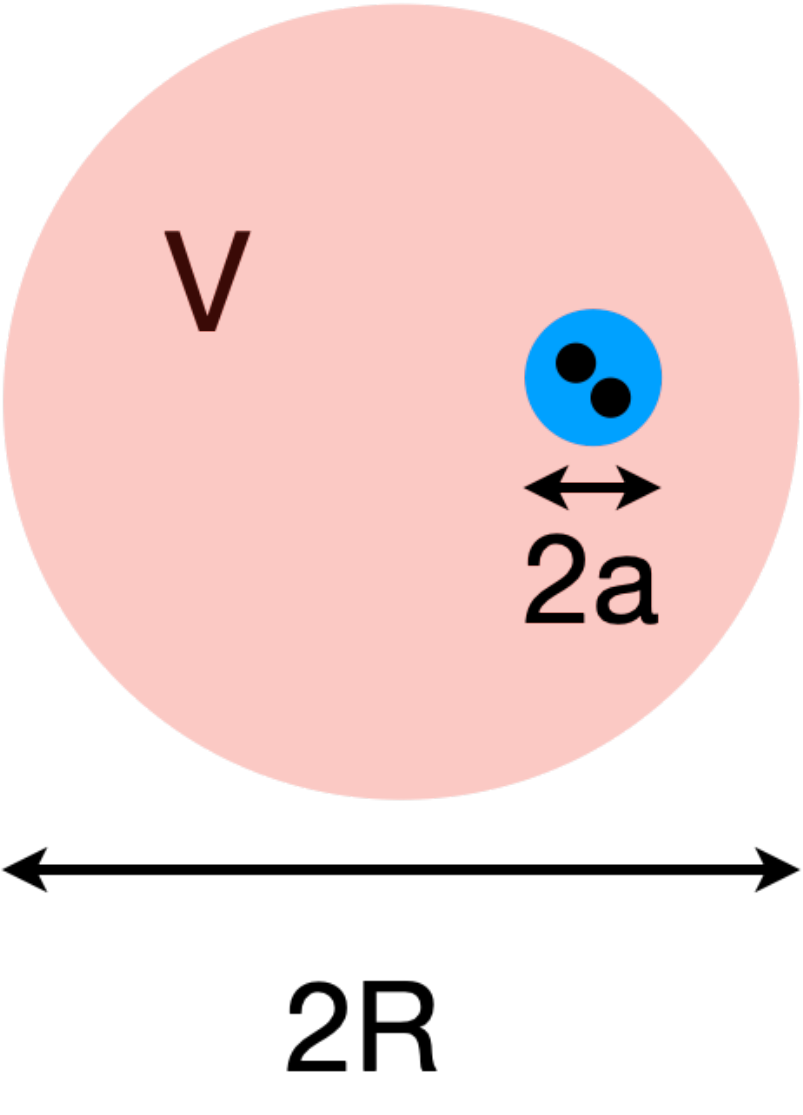}
\hspace{1cm}
\includegraphics[height=3cm]{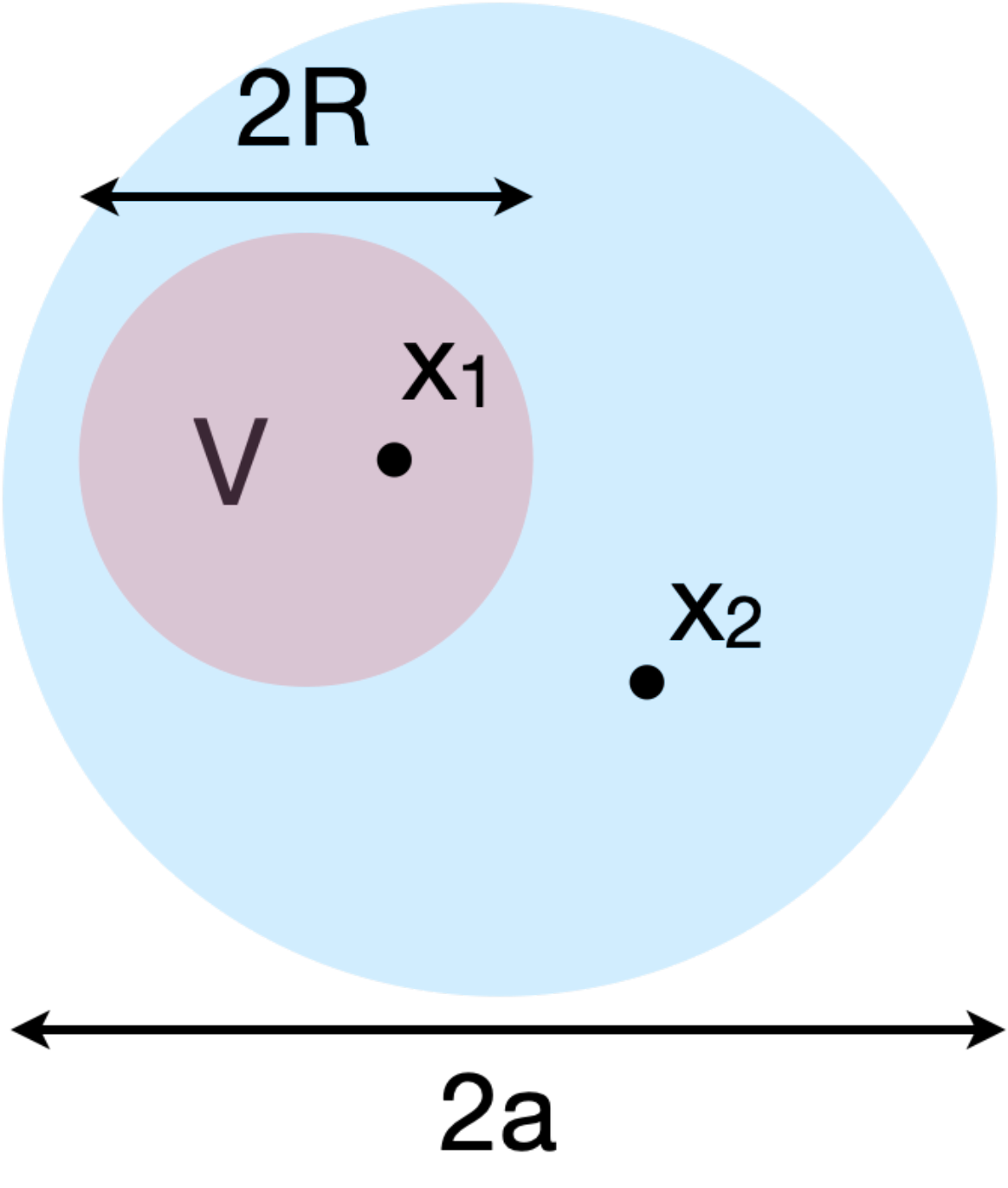}
\hspace{0.1\textwidth}
\includegraphics[height=4cm]{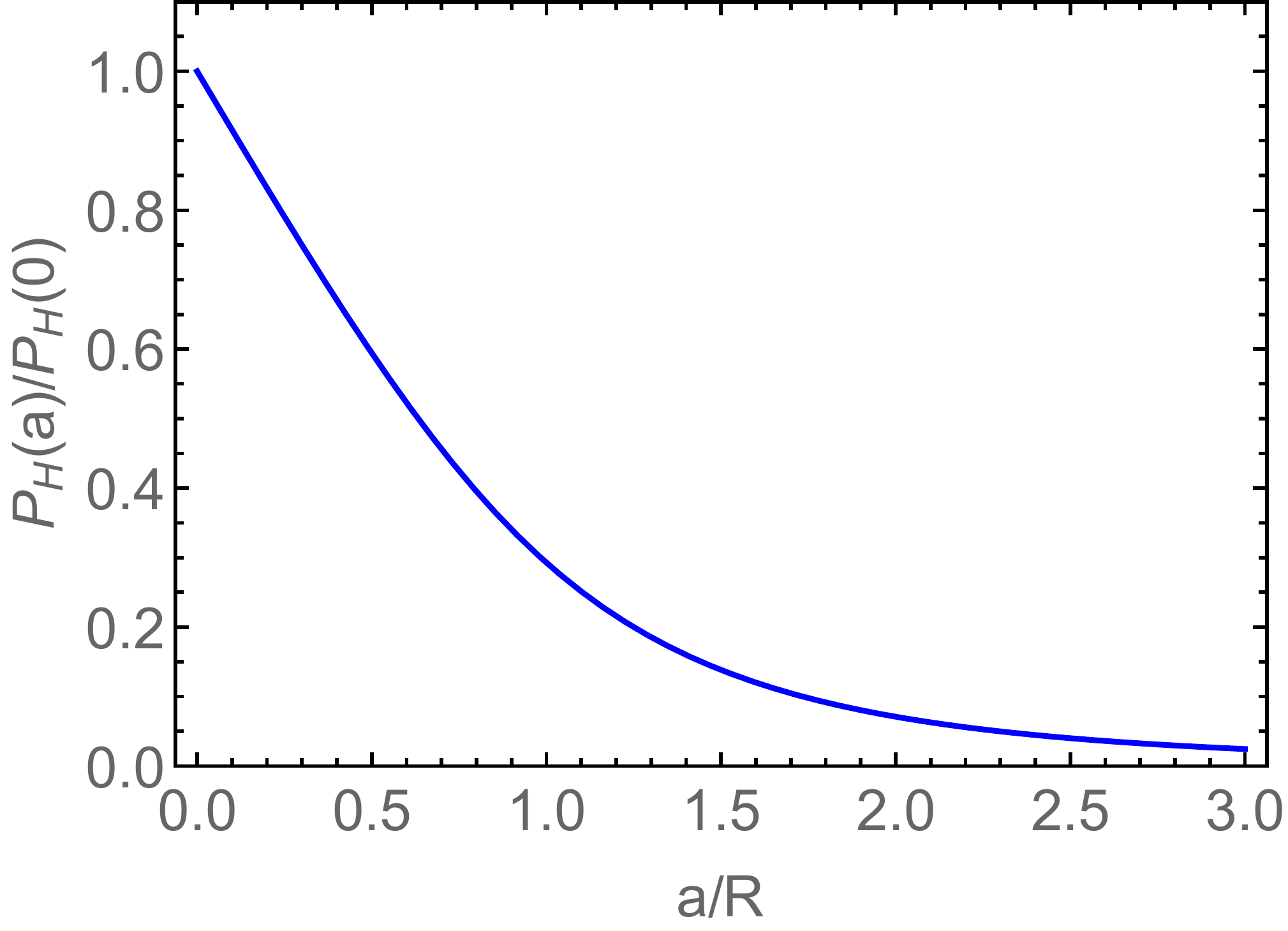}
\caption{Left panel: The emitted particle (radius $a$) is much smaller than the fireball (radius $R$). Middle panel: The emitted particle is much larger than the fireball. Right panel: Ratio of the particle yield calculated from Eq.~(\ref{eq-2}) for a Gaussian wave function as a function of the ratio $a/R$.}
\label{fig-1}
\end{figure}

I believe that this conclusion is erroneous, because the relative sizes of the emitting source and the emitted particle also comes into play when one performs a careful evaluation in the SHM. This is illustrated in Fig.~\ref{fig-1}. The left panel depicts the case when the emitted particle is much smaller   than the emission volume ($a \ll R$), which is typically the case for standard hadrons such as the $\Lambda$. The central panels depicts the opposite case  when the emitted particle is much larger than the emission volume ($a \gg R$), which is realized for small fireballs such as those formed in p+Pb and p+p and the $^3_\Lambda$H.

\begin{figure}[htb]
\centering
\sidecaption
\includegraphics[width=6cm]{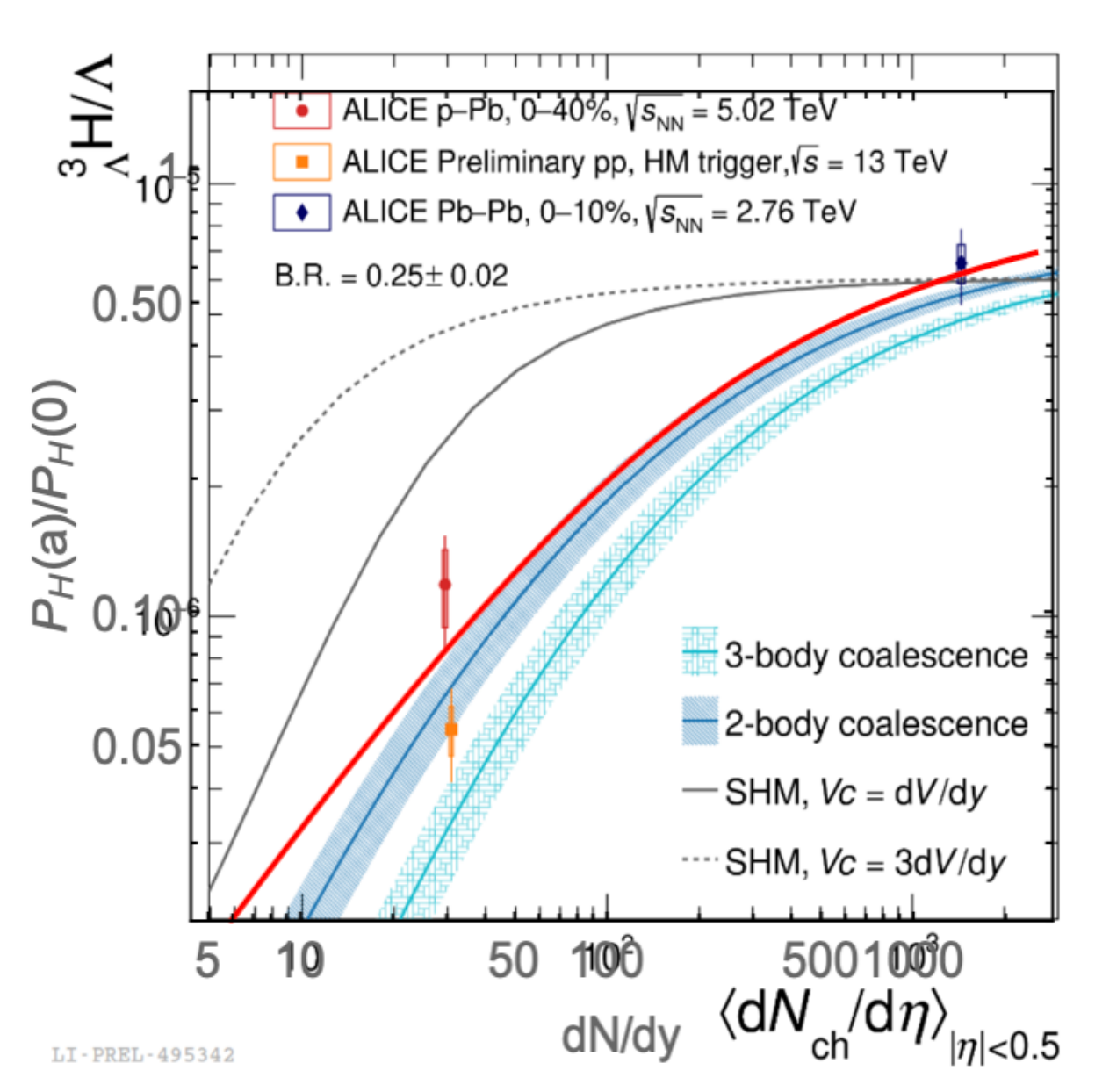}
\caption{Yield ratio $^3_\Lambda{\rm H}/\Lambda$ as a function of $dN_{\rm ch}/d\eta$ measured in Pb+Pb, p+Pb, and p+p collisions at LHC. The red curve shows the prediction for the SHM with size correction, Eq.~(\ref{eq-2}). The curves with uncertainty bands show the predictions of coalescence models. The solid and dotted curves show the predictions of the SHM with strangeness neutrality for two different values of the strangeness correlation volume $V_C$.}
\label{fig-2}
\end{figure}

When one calculates the yield of a hadron in the thermal model, one needs to evaluate the expression $P = {\rm Tr}[e^{-H/T}\,\theta_V]$, where $\theta_V=1$ inside the thermal volume and $\theta_V=0$ outside, assuming the fireball is surrounded by vacuum. One can then integrate out the internal wave function $\psi(x)$ of the emitted particle and obtains the usual SHM result:
\begin{equation}
P_0 \sim \int d^3p\, e^{-E_p/T} \int d^3x_1\, d^3x_2\, |\psi(x_1-x_2)|^2 \theta_V(x_1)\theta_V(x_2) 
\approx V \int d^3p\, e^{-E_p/T} .
\label{eq-1}
\end{equation}
On the other hand, when $a \gg R$ emission of the hadron requires that both (or generally all) constituents of the emitted particle to emerge from the hot fireball, and one finds:
\begin{equation}
P(a) \sim \int d^3p\, e^{-E_p/T} \int d^3x_1\, d^3x_2\, |\psi(0)|^2 \theta_V(x_1)\theta_V(x_2) 
\approx V^2 |\psi(0)|^2 \int d^3p\, e^{-E_p/T} .
\label{eq-2}
\end{equation}
In general, $P(a)/P_0$ is a function of the ratio $a/R$, which is one at $a/R=0$ and tends to zero for $a/R\to\infty$. Using a Gaussian internal wave function one obtains the plot shown in the right-most panel of Fig.~\ref{fig-1}. Inserting a reasonable estimate of $a=5$ fm for the radius of the hypertriton and scaling the radius $R$ of the fireball with the measured charged particle multiplicity $dN_{\rm ch}/d\eta$ one finds that the size-corrected prediction of the SHM closely tracks the prediction for two-body coalescence and agrees within uncertainties with the measurements for p+Pb and p+p, as shown in Fig.~\ref{fig-2}.

\section{Other topics}
\label{sec-4}

Other topics that were widely discussed in the conference were heavy quarkonia and strangeness. B.~Gossiaux \cite{Gossiaux} gave a thorough overview of the theoretical developments relating to quarkonium suppression. The main take-aways were: 
\begin{itemize}
\item The theoretical description of $\Upsilon$ suppression has reached a mature state with general agreement about the formal framework based on open quantum systems and NRQCD \cite{Rothkopf:2019ipj}. What is now needed are high statistics data. These will be forthcoming in the next few years from RHIC and LHC.  It remains to be seen whether the effect of color screening (Debye mass) on the data be isolated. $\Upsilon$ suppression at $p_T >$ few GeV requires more theoretical work (perhaps by adding SCET to the theoretical framework?).
\item Charmonium suppression is an extraordinarily complex process. Many mechanisms contribute:
nuclear shadowing (nPDFs) and other cold nuclear matter effects, color screening and thermal ionization; regeneration, especially at LHC, influenced by charm thermalization; and gluon fragmentation, which dominates at high $p_T$ and is affected by parton energy loss. A similarly comprehensive and rigorous theoretical framework  like the one for the $\Upsilon$ states does not yet exist for charmonium. The problem is that the charm quark mass is not large enough for a clean separation of scales.
\end{itemize}

There exists a veritable wealth of data on the production of strange hadrons; this area is not data limited as L.~Bianchi's \cite{Bianchi} survey talk made abundantly clear. It seems, however, that there has been only limited theoretical progress in recent years, possibly due to the success of the thermal resonance gas model. The effect of strangeness conservation in the fireball, often called ``canonical suppression'', is clearly observed STAR data at low collision energies ($\sqrt{s_{\rm NN}} = 2-3$ GeV) by looking e.~g., at the $\phi/K^-$ ratio.

Finally, there is continued progress in Bayesian model-to-data comparisons, which are becoming ever more ambitious. A new analysis by the Jyv\"askyl\"a group, presented by D.~J.~Kim \cite{Kim}, which used a much expanded set of observables, showed evidence of increasing discriminating power. These analyses require very substantial computing resources which will strain the resources available to the theoretical community. Clever approaches that reduce these requirements without sacrificing rigor are being explored.

\section{Final words}
\label{sec-5}

There are many other exciting developments which I have not covered. One example is the growing confluence between heavy ion collisions at lower energies, where the fireball contains a large surplus of baryons, and novel astrophysical observations for neutron stars (using $\gamma$-rays) and neutron star mergers (using gravitational waves). Both types of data have the ability to constrain the equation of state of cold, dense and hot, dense nuclear matter and connect these with analyses from heavy ion collisions. Clearly, the field remains vibrant and presents many opportunities for young scientists to make discoveries or to find answers to open questions. 

Last, but not least, I congratulate the SQM2022 organizers for conducting an inspiring conference that brought the interested scientific community together, either in person or remotely via video link. It was not at all easy to organize the conference under the conditions imposed by the pandemic, and we all owe them much gratitude for their efforts.

\end{document}